\documentclass[prb,showpacs,twocolumn,superscriptaddress]{revtex4}

\usepackage{graphicx}
\usepackage{amsmath}
\usepackage{amssymb}
\usepackage{epsfig}

\renewcommand{\v}[1]{{\bf #1}}

\newcommand{\w}{{\omega}}

\def\eqa{\begin{eqnarray}}
\def\eea{\end{eqnarray}}
\newcommand{\eq}{\begin{equation}}
\newcommand{\ee}{\end{equation}}

\newcommand{\<}{\langle}
\renewcommand{\>}{\rangle}

\newcommand{\p}{\partial}

\newcommand{\ua}{\uparrow}
\newcommand{\da}{\downarrow}
\newcommand{\ra}{\rightarrow}

\newcommand{\del}{\delta}

\newcommand{\La}{\Lambda}

\newcommand{\Om}{\Omega}

\newcommand{\si}{\sigma}

\usepackage{color}
\begin{document}

\title{Matrix element interference in $N$-patch functional renormalization group}

\author{Li-Han Chen}
\affiliation{National Laboratory of Solid State Microstructures \& School of Physics,
	Nanjing University, Nanjing, 210093, China}

\author{Zhen Liu}
\affiliation{National Laboratory of Solid State Microstructures \& School of Physics,
	Nanjing University, Nanjing, 210093, China}

\author{Jian-Ting Zheng}
\email{mf1722040@smail.nju.edu.cn}
\affiliation{National Laboratory of Solid State Microstructures \& School of Physics,
	Nanjing University, Nanjing, 210093, China}
\affiliation{Collaborative Innovation Center of Advanced Microstructures, Nanjing University, Nanjing 210093, China}

\begin{abstract}
	We show that the $N$-patch functional renormalization group (pFRG), a theoretical method commonly applied for correlated electron systems, is unable to implement consistently the matrix element interference arising from strong momentum dependence in the Bloch state contents or the interaction vertices. We show that such a deficit could lead to results incompatible with checkable limits of weak and strong coupling. We propose that the pFRG could be improved by a better account of momentum conservation.
\end{abstract}

%
%75.30.Fv  Spin-density waves
%74.20.Rp  Pairing symmetries (other than s-wave)
%74.20.-z  Theories and models of superconducting state
%71.27.+a  Strongly correlated electron systems; heavy fermions
%64.60.ae  Renormalization-group theory

\maketitle

\section{Introduction}

Since it was proposed by Wetterich in 1991,\cite{wetterich} the functional renormalization group (FRG) has been receiving increasing attention and application for correlated electron systems.\cite{rmp,ptp} Based on the generating functional of one-particle irreducible (1PI) vertices, the FRG provides exact flow equations of the 1PI vertex functions $V_n$ of arbitrary order $n$ versus a running parameter $\La$, which could be flexibly taken as the infrared cutoff scale in energy (through momentum) or frequency, or any other parameter that can be used to regulate the single-particle Green's function in such a way that the infrared limit is approached gradually.\cite{wetterich} The FRG may be considered as a continuous comparison of thought-models that differ in the infrared energy scale. As the genuine model is approached, the two-point vertex function $V_2$ provides exact description of the single-particle self-energy, and the four-point vertex function $V_4$ provides exact description of the effective interactions between quasi-particles. Higher-order vertex functions could also be described. In comparison, the more traditional Wilsonian RG (WRG) eliminates (or integrates out) higher-energy degrees of freedom gradually in favor of the effective action for lower-energy ones, followed by rescaling.\cite{shankar} As a result, the behaviors of the quasiparticles at higher energy scales are lost. In simple terms, the single-particle phase space captured in FRG and WRG is complementary: Starting from high energy scales, the phase space in FRG increases by lowering the infrared cutoff scale $\La$, which would be the ultraviolet limit in WRG (without rescaling). The two methods are closely related, however. For example, the four-point vertices $V_4$ in FRG at the scale $\La$ may be taken as the bare interactions for quasi-particles in WRG below $\La$.\cite{rmp} The wide use of Coulomb pseudo-potential in the traditional theory of superconductivity can be taken as an example of this correspondence in the simplest form (of a parameter instead of a function).

In application of FRG to interacting electrons, several truncations have to be made for practical purpose. First, the flow of $n$-point vertices is related to $n$- and $n+2$-point ones, and this chain has to be cutoffed at a suitable level. Usually, the cutoff is made at $n=4$, while vertices at $n>4$ are ignored. Then the flow equations become closed. The rationale behind this truncation is as follows. Consider the degrees of freedom around the Fermi surface (FS). It can be shown that in generic cases, all $n=4$ vertices are marginal in the RG sense, while their frequency dependence, as well as all higher-order vertices, are irrelevant. This implies that it is sufficient to keep vertices up to $n=4$ and ignore its frequency dependence. The RG argument is valid provided the normal state instability occurs at low energy scales, and as such the above truncation is applicable to weak or moderately coupled systems. In particular, this truncation is unable to address the interesting Mott limit. However, FRG at this level is already far beyond the ladder approximation and the random phase approximation, since the latter ones bias a particular scattering channel, while FRG is designed to treat all scattering channels on equal footing. Second, even if we truncate the flow equations at $n=4$, it is still challenging to parameterize the four-point vertex function, which we write as $V(\v k_1, \v k_2,\v k_3,\v k_4)\equiv V(\{\v k_i\})$ henceforth. Here $\v k_i$'s are momenta (out of which three are independent by momentum conservation), and the dependence on other internal degrees of freedom are suppressed (together with the ignored frequency dependence). In the so-called $N$-patch FRG (pFRG henceforth),\cite{rmp,ptp} the Brillouin zone is divided into $N$ patches, and the momentum dependence of $V$ is projected onto that on the patches where the momenta lie, respectively. By momentum conservation in the original vertex, the truncated vertex becomes a tensor of three discrete independent patch labels. This scheme makes calculation feasible, and has been applied widely in various condensed-matter systems,\cite{rmp,ptp} gaining considerable insights into the competing orders in correlated electron systems.

The purpose of this communication is to show that the pFRG breaks hermiticity of four-point vertex $V$ and breaks the momentum conservation in loop integrations using the truncated $V$. As such, it is unable to implement consistently the matrix element interference (MEI) arising from strong momentum dependence in the Bloch state contents or the interaction vertices, which occurs in some multi-band systems. We demonstrate by revisiting a case study in the literature that such a deficit could lead to results incompatible with checkable limits of weak and strong coupling. We point out that momentum conservation can be improved by dividing the patch into segments, providing a direction for the improvement of pFRG.

\section{Inherent inconsistency in pFRG}

To begin with, we consider a system with the single particle state $|\v r\>$ in real-space basis and eigenstate $|\v k\>$ in the momentum-space. Internal degrees of freedom, such as spin, orbital and sublattice are left implicit in $|\v r\>$, and so are the band label in $|\v k\>$. For the interest of pFRG, only the bands cut by the Fermi level are considered. Using the $|\v k\>$ basis is an advantage since the Green's function appearing in loop integration, such as shown in Fig.1(a), is diagonal. The interaction part of the Hamiltonian can be written as, in real space,
\eqa H_I = \sum_{\{\v r_i\}} \psi_{\v r_1}^\dag \psi_{\v r_2}^\dag U(\{\v r_i\})\psi_{\v r_3}\psi_{\v r_4},\eea
where $\psi_\v r$ is the fermion annihilation operator at $\v r$. We have absorbed a possible global symmetry factor, which reduces over counting and depends on whether $U$ is fully antisymmetrized. This factor should be singled out of $U$ in application but does not matter for our purpose. For local interactions, most of the elements are zero except for those with $\v r_i$'s close to each other. Notice that translation symmetry requires
\eqa  U(\{\v r_i\})=U(\{\v r_i-\v r_0\}) \eea
for any space displacement $\v r_0$. Transforming into the momentum space, we obtain
\eqa H_I=\frac{1}{\Om}\sum_{\{\v k_i\}}\psi_{\v k_1}^\dag \psi_{\v k_2}^\dag U(\{\v k_i\})\psi_{\v k_3}\psi_{\v k_4} \del_{\v k_1+\v k_2-\v k_3-\v k_4},\eea
where $\Om$ is the volume of the system, the last discrete delta function expresses crystal momentum conservation, and
\eqa  U(\{\v k_i\}) = \Om\sum_{\{\v r_i\}} \<\v k_1|\v r_1\>\<\v k_2|\v r_2\> U(\{\v r_i\}) \<\v r_3|\v k_3\>\<\v r_4|\v k_4\>, \label{eq:vk}\eea
in the convention $\<\v r|\v k\>=e^{i\v k\cdot\v r}/\sqrt{\Om}$ for the plane-wave part under box normalization. The initial condition for the four-point vertices is simply given by $V(\{\v k_i\}) = U(\{\v k_i\})$. Momentum dependence arises from nonlocal interactions even for a one-band system. More importantly, it would also arise even if the interaction is local in real space but the content of the Bloch state $|\v k\>$, such as the spin, orbital and sublattice, depend strongly on momentum. For example, the kagome lattice has three inequivalent sites within the unitcell. The Bloch state $|\v k\>$ varies in sublattice-contents significantly as the momentum $\v k$ varies in magnitude and direction.\cite{kiesel,thomale,jianxin,qhwang} We will come back to this point shortly. Finally, even if the initial vertex is trivial apart from momentum conservation, it can develop momentum dependence during FRG, since the flow of the vertex, $\p V/\p\La$, is a complicated quadratic functional of $V$ itself.\cite{wetterich,rmp,ptp} When the vertex diverges during the FRG flow, it is just the nature of the momentum dependence in it that determines how the normal state is going to develop instabilities. For example, if the eigenvalue of $V(\v k,-\v k,-\v p,\v p)$ (taken as a scattering matrix) diverges, it implies an emerging superconducting order in the particle-particle (pp) channel, while if the eigenvalue of $V(\v k+\v Q,\v p,\v k,\v p+\v Q)$ diverges, it implies a kind of density-wave order at momentum $\v Q$ in the particle-hole (ph) channel. Therefore, a proper account of the momentum dependence, referred to as MEI when it arises initially from the Bloch state,\cite{kiesel,thomale,jianxin,qhwang} is crucial to make sense of FRG.

Let us now discuss how the momentum dependence is treated in pFRG.\cite{rmp,ptp} For simplicity but without loss of generality, we consider a single band cut by the Fermi level. The momentum dependence in the four-point vertex function $V(\v k_1,\v k_2,\v k_3,\v k_4)$, such as one of the two vertices in Fig.1(a), is projected onto the dependence on the patch numbers, say $V(1,2,3,.)$,  whenever the momentum $\v k_i$ is in patch-$i$. See Fig.1(b) for illustration of the patches. Henceforth the dot argument in $V$ means the momentum passive to the other three active ones by momentum conservation. Apparently, by ignoring the radial dependence, momentum conservation is spoiled in the truncated vertices. Furthermore, in pFRG the momentum at FS is chosen as the representative one for a patch. This is used, for example, to calculate the initial vertex according to Eq.\ref{eq:vk}. However, even if the active momenta are set on the FS, in general the passive momentum can deviate from the FS. Take Fig.1(b) as an example: If $\v k_{1, 2, 4}$ are on the FS, $\v k_3$ is at the origin; while if $\v k_{1,2,3}$ are on the FS, $\v k_4$ wanders to a different patch far from the previous case. Therefore the patch parametrization would cause $V(1,2,3,.)$ to be different to $V(.,2,3,4)$, $V(1,.,3,4)$ and $V(1,2,.,4)$, if the active momenta are fixed at the FS. This can lead to violation of hermiticity in the resulting truncated vertices, since $V(4,3,2,.)\neq V^*(1,2,3,.)$ in generic cases. (The only exceptions are in scattering processes with zero Mandelstam momentum.) To be even worse, it is ambiguous in the loop integration, such as Fig.1(a). As $\v k_3$ is integrated within patch-3, Fig.1(b) shows that $\v k_4$ may trespass a significant fraction of the Brillouin zone, here chosen for a kagome lattice for later convenience. Keeping using the truncated $V(1,2,3,.)$ for the left vertex in Fig.1(a), without accounting for the patch variation of $\v k_4$, loses or smears the strong momentum dependence in the interaction vertex itself. A similar problem arises for the right vertex. This problem is quite general and can be even worse than that in Fig.1. Moreover it can not be fixed by increasing the patch number. The effect might be less severe if MEI is absent initially, as the application of pFRG in the square lattice seems to produce sensible results \cite{honerkamp} in good agreement with other FRG approaches,\cite{husemann} where momentum conservation is not violated but the fermion bilinears are limited in the various scattering channels.

\begin{figure}
	%\vspace{0.5cm}
	\includegraphics[width=\columnwidth]{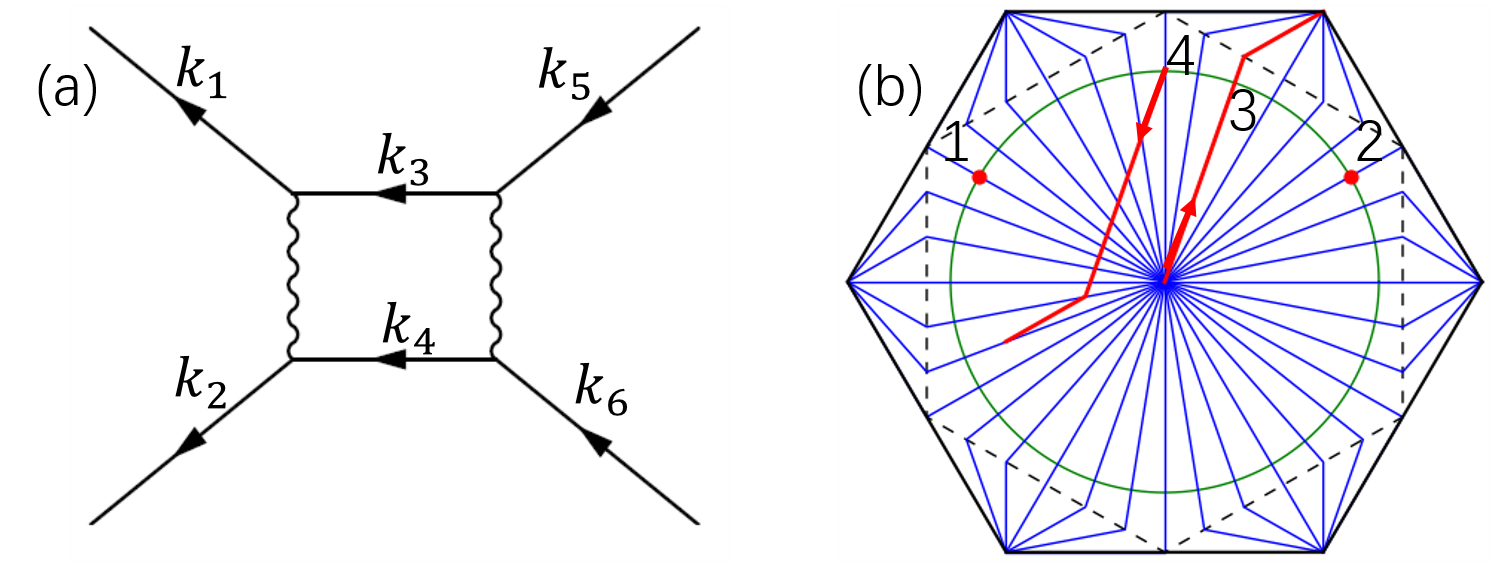}
	\caption{(a) A loop integration over momentum $\v k_3$ (and $\v k_4$ passively). The wavy lines are interaction vertices, and the straight lines are drawn for fermions. (b) Patch trespassing of $\v k_4$ as $\v k_3$ is integrated within patch-3. The external momenta $\v k_{1,2}$ are fixed at the red spots. The arrows show where $\v k_{3,4}$ start during the integration. The outer hexagon is the Brillouin zone boundary, and the green circle represents a generic Fermi surface, whose shape and location do not matter for the purpose of the illustration. }
\end{figure}

We should stress that the above inconsistency is present even for low energy quasiparticles sufficiently close to the Fermi surface. As we said above, in general it is impossible to set all of the four momenta of the interaction vertex on the Fermi surface. In fact it is because of this limitation that at low energy scales, only three types of vertices with zero Mandelstam momentum are left, namely, $V(\v k, -\v k, -\v p, \v p)$, $V(\v k, \v p, \v p, \v k)$, and $V(\v k, \v p, \v k, \v p)$, in the case of a generic Fermi surface. The flow of such vertices in Wilsonian RG is discussed in depth in Ref.\onlinecite{shankar}, and the validity is limited to weak coupling because of the restriction near the Fermi surface.   However, this is not the main point of the gist of FRG. In fact, the Wetterich equation, the basis of FRG, is exact,\cite{wetterich}  and hence is not limited to weak coupling in principle. Restricting to degrees of freedom on the Fermi surface loses much of the juices of FRG. Instead, the aim of FRG is to extract the effect of virtual excitations at higher energy scales on the effective interaction vertices, not only in strength but also in the functional form, for lower energy quasiparticles. To bring about such a nontrivial effect, it is important to treat degrees of freedom at all energy scales consistently, since the inconsistency at higher energy scales would lead to misleading or incorrect renormalized interaction vertices, and eventually lead to incorrect prediction of the emerging electronic order. Of course, in practical applications to electronic systems, the interaction vertices are truncated up to the forth order. This limits the regime of bare interactions in such a way that the instability of the normal state should occur only at low energy scales. Going beyond forth-order interaction vertices is interesting but beyond the scope of this work.

\section{Effect of the inherent inconsistency in pFRG: a case inspection}

Let us check some specific consequences of the inherent inconsistency in the implementation of pFRG, when MEI is important. Fortunately (or unfortunately), examples can be found easily in the literature. We consider the kagome-Hubbard model. As we mentioned above, the sublattice content of the Bloch state depends on momentum strongly.\cite{kiesel, thomale,jianxin,qhwang} This has profound consequences. For example, even though the FS is perfectly nested when the Fermi level is at the van Hove singularity (vHS), a local Hubbard interaction is inefficient in enhancing ph scattering at the nesting vector, since the latter connects equal-energy Bloch states of almost orthogonal sublattice contents. Instead, forward scattering is enhanced by the vHS. This type of MEI are reported to cause unusual Fermi-surface instabilities.\cite{kiesel,thomale,jianxin,qhwang} Specifically, using pFRG the authors in Ref.\onlinecite{thomale} reported persistence of ferromagnetic (FM) order for the Hubbard interaction $U$ up to ten times of the hopping $t$. This does not seem to fit well to two checkable limits of weak and strong coupling, as we discuss below.

\subsection{Stoner theory in the weak coupling limit}

\begin{figure}
	\includegraphics[width=\columnwidth]{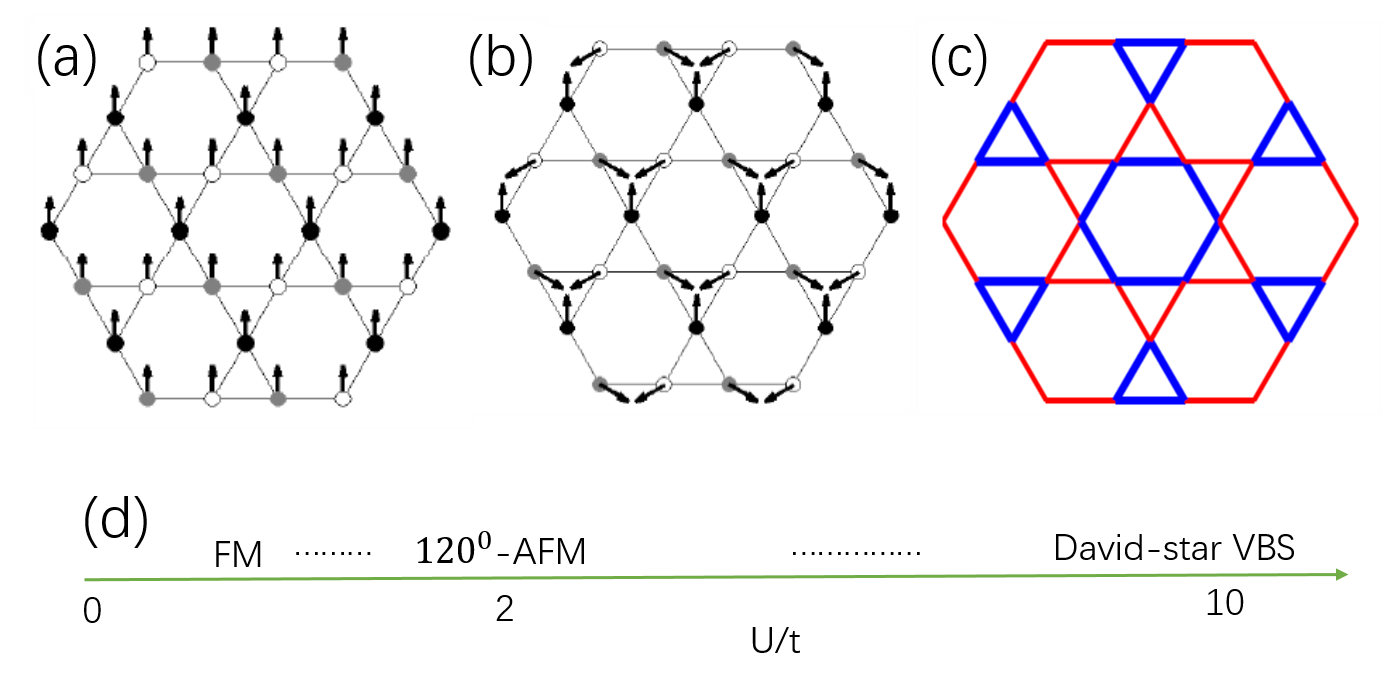}
	\caption{Illustration of (a) FM, (b) $120^\circ$-AFM, and (c) David-star valence-bond solid order on the kagome lattice. The arrows in (a) and (b) indicate the direction of the spin moment. The blue bonds in (c) form hexagons and triangles, and the blue/red bonds are associated with enhanced/weakened effective hopping. (d) A schematic phase diagram along the $U$-axis of the kagome-Hubbard model. The dots indicate possible transition between neighboring phases. }\label{fig:orders}
\end{figure}

In the weak coupling limit, we can apply the Stoner theory to determine the correct form of the spin order. We calculate the bare spin susceptibility $\chi_0(\v q)$, which is a function of momentum $\v q$ and a matrix in the sublattice basis, with the elements
\eqa \chi_0^{ab} = -\frac{T}{\Om}\sum_{\v k,\w_n} G_\v k^{ab}(i\w_n)G_{\v k+\v q}^{ba}(i\w_n), \eea
where $a$ and $b$ refers to sublattices, and $G_\v k(i\w_n)$ is the normal state Green's function (in the sublattice basis) at momentum $\v k$ and Matsubara frequency $\w_n$, for the tight-binding model on the kagome lattice. Notice that the MEI is encoded properly here in the sublattice dependence of the Green's function. As shown in Ref.\onlinecite{qhwang}, the leading eigenvalue of $\chi_0$ is largest at $\v q=0$ in the momentum space. So we will limit ourselves to $\v q=0$. For definiteness we denote the eigenvalue of $\chi_0$ as $\chi$. There are three eigenvalues, out of which one is nondegenerate, and the others two-fold degenerate. The former is associated with the eigenvector $(1,1,1)^t/\sqrt{3}$, corresponding to uniform ferromagnetic (FM) order within the unitcell. See Fig.\ref{fig:orders}(a) for illustration of this order. The other degenerate eigenvalues are associated with the eigenvectors $(1,-1,0)^t/\sqrt{2}$ and $(1, 1, -2)^t/\sqrt{6}$. It is known that this doublet actually recombines in the ordered state to form intra-unitcell antiferromagnetic order with an angle of $120^\circ$ between nearby spins ($120^\circ$-AFM henceforth).\cite{qhwang} See Fig.\ref{fig:orders}(b) for illustration of this order. Fig.\ref{fig:mft}(a) shows the Stoner instability lines $U=1/\chi$ for the two types of spin orders versus the temperature $T$. As $T$ is lowered, the first instability determines the type of spin order, or the mean field ground state. It is seen that the FM order is limited to $0<U<U_0$, where $U_0\sim 1.57t$. For $U>U_0$, the $120^\circ$-AFM is more favorable, exactly because of the MEI in $\chi_0$. Note the transition point $U_0$ is about a half of the sub-band cut by the Fermi level, or a quarter of the entire band structure, so the transition should be in or near the weak coupling limit, and hence should be trustable. On the other hand, the reason of the transition from FM to $120^\circ$-AFM is just because of the MEI emphasized so far. But the transition is not seen in Ref.\onlinecite{thomale}, although MEI was claimed to have been included in pFRG therein.

We are aware of the fact that the Stoner theory is equivalent to the Hatree-Fock mean field theory (HFMFT). The HFMFT is not exact, and neither is the truncated FRG. However, as far as the orders involving site-local spins are concerned, the HFMFT is actually quite suggestive. For example, the presence of 120$^\circ$-AFM order at the MF level corresponds to an attractive scattering channel, which is more favorable than FM at the tree level of FRG and hence is unlikely to disappear during FRG. On the other hand, the absence of FM at large $U$ in HFMFT acts strongly against its persistence in pFRG, since HFMFT has already over emphasized any order within its scope, and correlation of fluctuations beyond HFMFT, hopefully capturable by pFRG, could not possibly rescue the FM order. Therefore, Ref.\onlinecite{thomale} seems to have biased FM more than HFMFT would do, to the extent that the other and more favorable orders are overlooked at $U>U_0$, most likely because of the inherent inconsistency in pFRG discussed above.

\subsection{Dynamical mean field theory}
\begin{figure}
	\includegraphics[width=0.9\columnwidth]{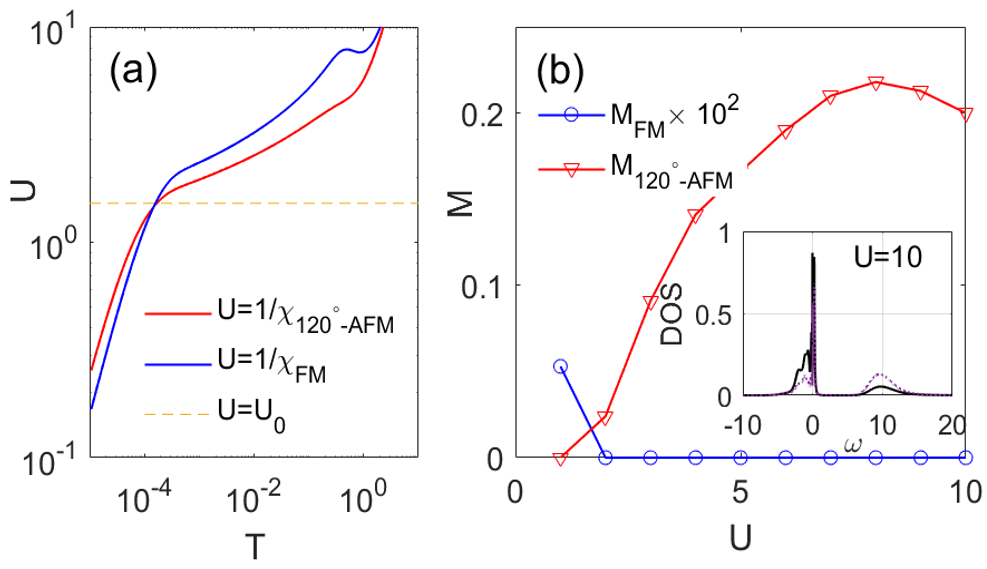}
	\caption{(a) The Stoner instability lines for FM (red) and $120^\circ$-AFM (blue) order in the kagome-Hubbard model. $U_0\sim 1.57$ is the crossing point. (b) The average moment from DMFT at zero temperature. The inset shows the spin-diagonal DOS. }\label{fig:mft}
\end{figure}

To see the robustness of the $120^\circ$-AFM above $U_0$, we go one step further by resorting to dynamical mean field theory (DMFT). \cite{dmft} The DMFT solves a single quantum impurity embedded in a dynamical environment, exactly and self-consistently, and is known to be much better than the HFMFT since local quantum fluctuations are taken into account. We solve the quantum impurity problem using the numerical renormalization group (NRG).\cite{nrg} For the uniform FM phase, we take one site as the impurity, and the rest the environment. For the $120^\circ$-AFM, the spin order is non-collinear. We need to take care of three inequivalent spin moments. However, assuming the self-energy is local, we verified that the frequency-dependent self-energy matrix (in spin basis) on one site can be consistently rotated to that on the other sites by $\pm 120^\circ$ SU(2) rotations, according to the pattern shown in Fig.\ref{fig:orders}(b). Therefore, it is still possible to take one site as the quantum impurity (for which we take the spin-diagonal basis), and the rest as the environment, with the above necessary rotations in mind to form the lattice Green's function, or in the Hilbert transformation to get the hybridization kernel. A similar but simpler case is the AFM on square lattice, where there are two inequivalent spin sites. The subsequent calculations are standard. We start from a self-energy capable of inducing the respective form of spin order, and iterate until the hybridization kernel or the self-energy converges. The chemical potential is tuned concurrently to fix the local charge density at $5/6$.
Our DMFT results (for discrete values of $U$) are presented in Fig.\ref{fig:mft}(b), showing FM can not survive at $U\geq 2t$ while $120^\circ$-AFM is stable. This is consistent with Stoner theory (or HFMFT) qualitatively. But quantitatively, the size of the moment is reduced by local quantum fluctuations captured by DMFT. For example, at large $U$, the spin moment $M\ra 1/2$ in HFMFT, but $M<1/4$ in DMFT and begins to drop for $U>8t$.

\subsection{Strong coupling limit}

The HFMFT and DMFT discussed above are limited to ordering of local spin moments. The spatial correlation between fluctuations are neglected in both HFMFT and DMFT, but could wipe out the spin orders, in favor of ordering on bonds, such as the valence-bond solid (VBS). We notice that at large $U$, the Mottness comes into play, see the lower and upper Hubbard bands separated by a Mott gap in the inset of Fig.\ref{fig:mft}(b) calculated by DMFT for $U=10$.
At this stage, a better account of the Mott gap, short-range spin-spin correlation and renormalization of kinetic energy at the Hamiltonian level, is taken into account by the $t$-$J$ model,
\eqa H_{t-J}=-t\sum_{\<ij\>\si} (\tilde{c}_{i\si}^\dag \tilde{c}_{j\si} + {\rm h.c.}) + J\sum_{\<ij\>}(\v S_i\cdot \v S_j-n_i n_j/4),\nonumber\eea
where $\tilde{c}_{i\si}=P c_{i\si} P$, with $P=\Pi_i (1-n_{i\ua}n_{i\da})$, describes electrons in the lower Hubbard band,  $J=4t^2/U$, and $\v S_i$ is the electron spin. Systematic results near half filling of the kagome model are known in the literature,\cite{sigi11,sigi13,sigi14} where the David-star VBS, as illustrated in Fig.\ref{fig:orders}(c), is reported to be energetically favorable, and is more robust for larger $J$ (or lower $U$ in the Hubbard model but still in the doped Mott limit).

Combining our HFMFT and DMFT results as well as the VQMC results in the literature,\cite{sigi11,sigi13,sigi14} we may draw a schematic phase diagram for the kagome-Hubbard model, see Fig.\ref{fig:orders}(d). Qualitatively, the electronic order changes from FM at weaker $U$, to $120^\circ$-AFM for intermediate $U$, and to the David-star VBS in the strong coupling limit. This clearly falsifies the persistence of FM from small $U$ up to $U=12t$ obtained by the pFRG, \cite{thomale} showing the inherent inconsistency in pFRG is detrimental (at least) for this model. 

Several remarks for the phase diagram are in order. We should mention that the FM and AFM states are defined either at zero temperature or in the mean field sense at finite temperatures, since a continuous symmetry can not be broken at finite temperature by Mermin-Wagner theorem. The VBS is rather different. It only breaks discrete symmetry, and hence can order in two-dimension even at finite temperatures. In fact the VBS state generates a full gap for quasiparticle excitations, and should be much more stable than the other states, once established.

\section{Discussion and conclusion}

Having envisioned the inherent inconsistency in pFRG and its consequence in application, we ask how pFRG could be improved. We notice that the difficulty arises from the violation of momentum conservation in the truncated vertices. This is the key for improvement. A possible direction is to resolve the dependence along the radial direction of the patch. Ideally a full account of radial dependence would eliminate the inconsistency completely. In practice, a continuous resolution of the radial dependence is impossible. However, it is possible to divide a patch into several segments, and the vertex function could be parameterized as $V(1a,2b,3c,.)$ where $(1,2,3)$ label the patches and $(a,b,c)$ label the segments within the patches. The dotted argument will have slaved patch and segment labels. The segments should be chosen in such a way that integration within one segment in one patch, say that for $\v k_3$ in patch-3, does not lead to change of the patch and segment label for the slaved momentum, say $\v k_4$, or at least the change does not occur frequently. In this way, the inconsistency of pFRG could be weakened. Work along this direction is underway.

The other direction is to view the four-point vertices as scattering matrices for fermion bilinears in various channels, and subsequently truncate the internal spatial range within the bilinears. Momentum conservation can then be implemented exactly, so that MEI is captured without difficulty. In an early attempt, the bilinears are limited to local spin-density in the ph channel and $d$-wave Cooper pairing on first-neighbor bonds in the pp channel.\cite{husemann} This simple truncation already proves successful in the Hubbard model for cuprates. The idea has been systematically extended to include all possible short-range bilinears in both pp and ph channels on equal footing, up to a cutoff length that is sufficient to describe potentially singular scattering modes that provide the microscopic structures of the emerging orders.\cite{tSC,graphene} The corresponding FRG scheme is called singular-mode (SM) FRG and has been applied in various contexts.\cite{qhwang,tSC,graphene,biaxial,bc3,LaOCaAs,uniaxial,kagome-sd2,FeS,holstein,loop-current,SrPtAs,SrIrO,epl,FeSe,BiS,FeSe-SrTiO} Note that the parametrization of four-point vertices is asymptotically exact in SM-FRG as the truncation length increases. In contrast, the truncation error always exists in pFRG due to the ignored radial dependence along the patches, and the inherent inconsistency regarding MEI can not be eliminated by just increasing the patch number. Indeed, both the $120^\circ$-AFM order and the VBS order are discovered by SM-FRG in Ref.\onlinecite{qhwang} as the bare interaction $U$ increases, in contrast to the persistence of FM in pFRG.\cite{thomale}. This is a clear indication of the importance of momentum conservation and hermiticity, which are unfortunately violated in pFRG.

To conclude, we have demonstrated that in pFRG the strong momentum dependence in the vertex, or the matrix element interference, can not be implemented consistently. This could be a severe problem in systems where MEI matters through strong momentum-dependence in Bloch state contents and/or bare interactions.\cite{nacoo} In fact, since the vertex function develops strong momentum dependence during FRG even if it is initially trivial, the interpretation of the pFRG results for simple systems should also be taken with some grains of salt. We proposed a direction for the improvement of pFRG via better resolution of momentum conservation, and we found that the SM-FRG is in fact superior in comparison.\\

\acknowledgments{We thank Ronny Thomale for drawing our attention to the matrix element interference and encouraging us to take a deep look into the functional renormalization group. The project was supported by the National Key Research and Development Program of China (under Grant No. 2016YFA0300401), and the National Natural Science Foundation of China (under Grant No.11574134).}

\end{document}